# Flexomagnetoelectric Interaction in Cubic, Tetragonal and Orthorhombic Crystals

B. M. Tanygin[a]

[a] Radiophysics Department, Taras Shevchenko Kyiv National University, 4G, Acad. Glushkov Ave., Kyiv, Ukraine, UA-03127

*Corresponding author:* B.M. Tanygin, Radiophysics Department, Taras Shevchenko Kyiv National University, 4G, Acad. Glushkov Ave., Kyiv, Ukraine, UA-03127.

*E-mail*: b.m.tanygin@gmail.com

*Phone*: +380-68-394-05-52

**Abstract.**

The phenomenological theory of the flexomagnetoelectric coupling in crystals of the cubic, tetragonal and orthorhombic crystal systems has been suggested. Secondary role of the crystal structure chirality was shown. Oppositely, significant role of the crystallographic point group type (symmetric, alternating, dihedral or cyclic) in the flexomagnetoelectric coupling has been derived. It was shown, that conceptually new features of the flexomagnetoelectric effects are expected in the crystals of the cyclic groups (crystal classes $4/m$, $\bar{4}$ and 4). Proposed verification of the theory is investigation of the domain wall bend details (changes of the effect symmetry). Special case of such verification near the compensation point is suggested. First-principles mechanisms of the flexomagnetoelectric interaction were discussed.

## 1. Introduction

Coupling mechanism between magnetic and electric subsystem in magnetoelectric materials [1] is of considerable interest to fundamentals of condensed matter physics and to the applications in the novel multifunctional devices [2] including low-power-consumption spintronic and magnonic devices. The magnetoelectric coupling is possible between homogeneous magnetization and electric polarization [3-7], and, also, between inhomogeneous magnetization and polarization [8-27]. Latter type of coupling was called as flexomagnetoelectric (FME) interactions [15-22]. Hypothetical coupling with inhomogeneous polarization has been formulated on the basis of the symmetry analysis [23]. The FME coupling is possible not only between existing magnetic and ferroelectric orderings (i.e. in the multiferroic material). It, also, describes inducing of the polarization in the region of the magnetic inhomogeneity [14]. The FME effects have been observed as the motion of magnetic domain walls in the gradient electric field in $(BiLu)_3(FeGa)_5O_{12}$ films [25,26]. The FME interaction was described by the four-constant coupling term in the cubic $m\bar{3}m$ crystal [28]. The phenomenological theory of the FME coupling was not yet created for other symmetry classes, including cubic system classes (classes 432, $\bar{4}3m$, $m\bar{3}$ and 23).

The purpose of this work is building of the phenomenological theory and systematization of the corresponding microscopic mechanisms of the FME interactions in the cubic, tetragonal and orthorhombic magnetically ordered crystals.

## 2. Microscopic origin of the flexomagnetoelectric coupling in ferri- and ferromagnetic crystals

### 2.1 Phenomenological statement of problem

Experimental verification of the FME coupling was provided [15-22]. Symmetry based analysis allows this interaction in any magnetically ordered medium [29]. However, microscopic mechanism of the FME coupling still requires attention due to the variety of possible phenomena in different ordered media. Systematization of these mechanisms is suggested in the present work. The application of the approach to the cubic, tetragonal and orthorhombic magnetically ordered crystals is only constructive example. However, these crystals families cover variety of the ferri- and ferromagnetic crystals, which are studying in the condensed matter physics.

The form of the coupling of the electric polarization **P** to the magnetization **M** can be obtained via the general symmetry arguments [30,31]. Such contribution to the free energy density is given by the expression [13] (here and hereinafter, we use Einstein notation):

$$F_{ME}(\mathbf{r}) = \iint D_{ijn}(\mathbf{r}-\mathbf{r}',\mathbf{r}-\mathbf{r}'') P_i(\mathbf{r})M_j(\mathbf{r}')M_n(\mathbf{r}'')d^3\mathbf{r}'d^3\mathbf{r}'', \quad (1)$$

where integration is performed over the volume of the crystal. The expression (1) describes both local (radius of interaction $a$ is determined by size of the magnetic lattice unit cell) and non-local (radius of interaction is determined by the crystal sample size) interactions [8,13]. The simplest case of the latter interaction is magnetoelectric coupling via (inverse) magnetostrictive and (inverse) electrostrictive interactions [1,23]. Indirect verification of this schema was provided [32-34]. The theory and comparison with experiment in case of the epitaxial ferroelectric and magnetic nanocomposite thin films were suggested [35].

The series expansion of (1) in terms of the small interaction radius gives free energy of the local FME interactions [8]:

$$F_{ME} = \gamma_{ijkl} P_i M_j \nabla_k M_l \quad (2)$$

The structure of the tensor $\hat{\gamma}$ and, consequently, the whole expression (2) is determined by the magnetic point group (crystallographic point group) of the crystal in the paramagnetic and paraelectric phase $G_P$ [8,13], i.e. (2) must be the invariant of this group. Only the local FME interactions of the form (2) are considered in scope of the present work.

**2.2 Relation between symmetry based phenomenology and microscopic mechanism**

Microscopic origin of different terms in the sum (2) was classified by Bar'yakhtar [13] as: exchange, exchange-relativistic and relativistic magnetoelectric coupling. According to this classification, the most significant contribution to the FME coupling is provided by the exchange term:

$$F_1 = \lambda_{ik} P_i \nabla_k \mathbf{M}^2, \tag{3}$$

where tensor $\hat{\lambda}$ (and all further tensors) is derived from the $\hat{\gamma}$. The next in order of magnitude term is the antisymmetric exchange-relativistic Lifshitz invariant [13]:

$$F_2 = d_{ijkl} P_i (M_j \nabla_k M_l - M_l \nabla_k M_j), \tag{4}$$

where $\hat{d}$ is antisymmetric with respect to a transposition of indices $j$ and $l$. The symmetric relativistic invariants have smallest magnitude [13]:

$$F_3 = \beta_{ijkl} P_i \nabla_k (M_j M_l), \tag{5}$$

where $\hat{\beta}$ is symmetric with respect to a transposition of indices $j$ and $l$.

The significance of the magnetization total derivatives terms (3) and (5) in the magnetoelectric material was recently discussed [28]. A novel route for the potential strong coupling multiferroicity was discussed in the Ref [36], where important role of the term (3) was shown. This term affects magnetization and/or polarization distribution only when the magnetization magnitude becomes spatially modulated. The example is the acentric dislocated spin-density wave ordering. On the basis of the experimental results, it was found, that this mechanism is substantial in the $YMn_2O_5$ [36].

Also, the well investigated microscopic mechanism exists for the invariant (4). Microscopic nature of the exchange-relativistic FME interaction has been suggested based on the spin supercurrent model with the same expression for cases (different $d$) of the superexchange and double-exchange interactions [37]:

$$\mathbf{P}_{ME} = -d\mathbf{e}_{12} \times [\mathbf{S}_1 \times \mathbf{S}_2], \tag{6}$$

where $\mathbf{S}_1$ and $\mathbf{S}_2$ are spins of the transition metal ions in the $M - O - M$ three-atom triad. The $\mathbf{e}_{12}$ is a polar time-even [4] vector connected transition metal ions. It was suggested that the Dzyaloshinskii-Moriya (DM) interaction induces the polarization of the electronic orbitals, without the involvement of the lattice degrees of freedom [37]. Superexchange-driven vortex magnetoelectricity mechanism has been suggested [38]. More general theoretical analysis of the microscopic theory was presented in Ref. [39]. It was concluded that coupling between the inhomogeneous magnetization and the polarization is created by a generalized version of DM interaction in the environment of certain spin orders. It was shown [39], that bond-bending existing in transition metal oxides enhances ferroelectricity. The similar theory was suggested in case of the $\text{Mn} - \text{O} - \text{Mn}$ bonding in the perovskite multiferroic $R\text{MnO}_3$, ($R = \text{Gd}, \text{Tb}, \text{Dy}$) [40]. Similarly, the DM interaction induces the ferroelectric state through lattice deformations, and stabilizes the magnetic structure at low temperatures [40]. Symmetry based condition of existing of the DM interaction was formulated [41-44]. However, the above-mentioned generalized "magnetoelectric" version [37, 39, 40] of the DM interaction requires special symmetry analysis, which was described in the Ref. [43]. The microscopic spin-dependent polarization can be induced not only by the DM interaction. More general form was obtained [43,13]:

$$\mathbf{P}_{ME} = \lambda(\mathbf{S_1 S_2}) + \hat{d}[\mathbf{S_1} \times \mathbf{S_2}] + \mathbf{S_1}\hat{\Gamma}\mathbf{S_2}, \tag{7}$$

The investigations [37, 39] describe the second terms of (7). All terms of the (7) are related to the macroscopic expressions (3-5) [13].

Microscopic nature of the local FME interaction can have magnetoelastic origin [13]. Also, Ruderman-Kittel-Kasuya-Yosida interaction can enhance FME coupling. It was shown [45], that anisotropic term of the exchange interaction which is linear in the spin-orbit coupling can exist. Non-collinear spin structures can be induced by the next-nearest-neighbor hopping [46]. A generic mechanism that couples electric field to spin patterns in the Mott insulating ferromagnets independently of their crystal structure was suggested [47].

Generally, symmetry considerations capture any microscopic mechanism for the coupling between ferroelectricity and inhomogeneous magnetization [48]. Specific form of the material constant tensors in the expressions (3-5) will be obtained in the next section for crystals with different crystal classes.

## 3. Phenomenological theory in cubic, tetragonal and orthorhombic crystal systems

It is necessary to consider two separate cases when Lifshitz invariants appear in (2). Usually, they appear after removing of the total derivatives. Alternatively, antisymmetric Lifshitz invariant term can be required by the crystal symmetry. In Ref. [28] it was shown that total derivatives (3) and (5) are special types of the FME coupling and they could not be removed in case of the most general variation problem, which requires consideration of the unknown functions $\mathbf{M}(\mathbf{r})$ and $\mathbf{P}(\mathbf{r})$. In the present work, we use the same group-theoretical analysis as in the Ref. [28]. The generalization of the polarization inducing scalar form [14] to the tensor form is required in case of the uniaxial and biaxial crystals:

$$\mathbf{P}_{ME} = -\hat{\chi}_e \, (\partial F_{ME}/\partial \mathbf{P}) \tag{8}$$

The structure of energy (2) was obtained for all possible cubic, tetragonal and orthorhombic ferri- and ferromagnetic crystals (table 1). The method is the standard construction of the invariant using irreducible representations of the crystallographic point group of the crystal in the paramagnetic phase [8,13,28]. If taken into account that $\mathbf{M}$ is an axial time-odd vector, the $\mathbf{P}$ is a polar time-event vector, the action of operator $\nabla_k$ is the analog of the multiplication on the $k$-th component of the polar time-even vector then it is possible to find invariants of the type (2), which are allowed by the magnetic symmetry group $G_P$ of the crystal in the paramagnetic phase (e.g., group $m\bar{3}m1'$ in case of $m\bar{3}m$ cubic crystal, where $1'$ is a time reversal transformation). Each summand of the $F_{ME}$ expression (2) must either remain constant or transform into another summand after action of any symmetry operation of the group $G_P$. Otherwise, the summand $P_i M_j \nabla_k M_l$ must conforms to the $\gamma_{ijkl} \equiv 0$.

Renormalization of constants was performed in order to simplify expression in same manner as in Ref. [28]. Also, in order to simplify the form, the final invariants were not expressed into the traditional sum of the symmetric (5) and antisymmetric (4) terms. Symmetry groups are classified as: symmetric ($S$), alternating ($A$), dihedral ($D$) and cyclic ($C$) group [49]. These pure algebraic properties of the group of any kind directly determine the type of the FME invariants: $F_{ME}^S$ (4 phenomenological constants [28]), $F_{ME}^A$ (7 constants), $F_{ME}^D$ (11 and 21 constants in tetragonal and orthorhombic systems respectively) and $F_{ME}^C$ (21 constants) respectively. The invariant $F_{ME}^D$ of the orthorhombic crystals contains 4 constant tensors.

The centrosymmetric, polar and enantiomorphic types of point group do not correlate with the form of the FME energy invariants. Consequently, the FME coupling is not related to the enantiomorphism of the magnetically ordered media [50] in the paramagnetic phase, which conforms to conclusion [29] that the FME coupling is allowed in any magnetically ordered crystal (centrosymmetric, enantiomorphic or polar).

**Table 1.** Free energy of the flexomagnetoelectric coupling in the cubic, tetragonal and orthorhombic ferri- and ferromagnetic crystals

| Crystallographic group[a] | $F_{ME}$ [b] |
|---|---|
| **Cubic system** | $F_{ME}^S = \tilde{\gamma}_1 P_i \nabla_i M_i^2 + \tilde{\gamma}_2 (\mathbf{P}\nabla)\mathbf{M}^2 + \tilde{\gamma}_3 \mathbf{P}(\mathbf{M}\nabla)\mathbf{M} + \tilde{\gamma}_4 (\mathbf{PM})(\nabla\mathbf{M})$ |
| *Symmetric groups* $m\bar{3}m, 432, \bar{4}3m$ | |
| *Alternating groups* $m\bar{3}, 23$ | $F_{ME}^A = \tilde{\gamma}_1 P_i \nabla_i M_i^2 + (\tilde{\gamma}_2' + \tilde{\gamma}_2'' \varepsilon_{ijk}) P_j \nabla_j M_k^2 +$ $+ (\tilde{\gamma}_3' + \tilde{\gamma}_3'' \varepsilon_{ijk}) P_j M_k \nabla_k M_j + (\tilde{\gamma}_4' + \tilde{\gamma}_4'' \varepsilon_{ijk}) P_j M_j \nabla_k M_k$ |
| **Tetragonal system** | |
| *Dihedral groups* $4_z/m_z m_x m_{xy}$, $\bar{4}_z 2_x m_{xy}, 4_z m_x m_y$, $4_z 2_x 2_y$ | $F_{ME}^D = (\tilde{\gamma}_1' + \tilde{\gamma}_1'' \delta_{iz}) P_i \nabla_i M_i^2 + (\tilde{\gamma}_2' + \tilde{\gamma}_2'' \delta_{iz} + \tilde{\gamma}_2''' \delta_{jz}) P_i \nabla_i M_j^2 +$ $+ (\tilde{\gamma}_3' + \tilde{\gamma}_3'' \delta_{iz} + \tilde{\gamma}_3''' \delta_{jz}) P_i M_j \nabla_j M_i + (\tilde{\gamma}_4' + \tilde{\gamma}_4'' \delta_{iz} + \tilde{\gamma}_4''' \delta_{jz}) P_i M_i \nabla_j M_j$ |
| *Cyclic groups* $4_z/m_z, \bar{4}_z, 4_z$ | $F_{ME}^C = F_{ME}^D +$ $+ (1 - \delta_{iz} - \delta_{jz}) \varepsilon_{ijk} \times$ $\times (\tilde{\gamma}_5 P_i M_i \nabla_i M_j + \tilde{\gamma}_6 P_i \nabla_j M_i^2 + \tilde{\gamma}_7 P_i M_j \nabla_i M_i + \tilde{\gamma}_8 P_j \nabla_i M_i^2) +$ $+ \tilde{\gamma}_9 P_z M_z (\nabla_x M_y - \nabla_y M_x) + \tilde{\gamma}_{10} P_z (M_x \nabla_z M_y - M_y \nabla_z M_x) +$ $\tilde{\gamma}_{11} P_z (M_x \nabla_y - M_y \nabla_x) M_z + \tilde{\gamma}_{12} M_z (P_x \nabla_z M_y - P_y \nabla_z M_x) +$ $+ \tilde{\gamma}_{13} (P_x \nabla_y - P_y \nabla_x) M_z^2 + \tilde{\gamma}_{14} (P_x M_y - P_y M_x) \nabla_z M_z$ |
| **Orthorhombic system** [c] | |
| *Dihedral groups* $mmm, mm2, 222$ | $F_{ME}^D = \Gamma_{1ii} P_i \nabla_i M_i^2 + \Gamma_{2ij} P_i \nabla_i M_j^2 + \Gamma_{3ij} P_i M_j \nabla_j M_i + \Gamma_{4ij} P_i M_i \nabla_j M_j$ |

[a] Crystallographic point group of the simultaneous paramagnetic and paraelectric phase of the crystal

[b] $\varepsilon_{ijk}$ and $\delta_{iz}$ is the Levi-Civita symbol and Kronecker delta respectively

[c] Components $\Gamma_{\xi ii}$ are different material constants. Tensors $\hat{\Gamma}_\xi$ have no symmetries.

**4. Discussion**

The suggested phenomenological expressions can be simply generalized to case of the antiferromagnetic ordering as it was shown for the $Cr_2BeO_4$ (two antiferromagnetic vectors) [24] and multiferroic $BiFeO_3$ (single antiferromagnetic vector) [15,18].

The term (3) is substantial in the $YMn_2O_5$ [36]. In case of the ferrimagnets, this part of interaction can be experimentally investigated in the vicinity of the compensation point. Also, the strong enough electric field can produce large polarization which can change magnetization magnitude in the local region. Such concept can be utilized experimentally in order to detect and/or measure constant of the coupling via the (3) term in the given crystal. According to the present phenomenological theory, the term $\tilde{\gamma}_2(\mathbf{P}\nabla)\mathbf{M}^2$ exists only in the $m\bar{3}m$ crystal (table 1). Orthorhombic crystal $YMn_2O_5$ has the similar term $\Gamma_{2ij}P_i\nabla_i M_j^2$, which induced the similar mean polarization distribution. Basically, as a rule, all FME features of the $m\bar{3}m$ crystal remain in case of other cubic, tetragonal and orthorhombic crystals. Crystallographic point groups of these crystals are subgroups of the $m\bar{3}m$. Consequently, additional components of the induced polarization occur in the similar FME phenomena. Thus, in order to bundle the free energy invariant, which are responsible for the similar FME phenomena, the numbering of four phenomenological constant was retained for all crystal systems (table 1). New FME properties appear in case of the cyclic groups ($4_z/m_z$, $\bar{4}_z$, $4_z$) of the tetragonal crystal system, because additional types of the phenomenological terms occur, including Lifshitz invariant terms. Thus, the FME coupling in the crystals with cyclic crystallographic point group requires attention in the experimental research due to possibility of the detection of new magnetoelectric properties, which can be described by additional 10 constants ($\tilde{\gamma}_5$ - $\tilde{\gamma}_{14}$) in the tetragonal crystal system.

The terms $(1-\delta_{iz}-\delta_{jz})\varepsilon_{ijk}\tilde{\gamma}_8 P_j\nabla_i M_i^2$ and $\tilde{\gamma}_{14}(P_x M_y - P_y M_x)\nabla_z M_z$ produce similar phenomenon in the same experimental setup, as described in Ref. [26], where out-of-film polarization was induced. Difference is the inducing of the unidirectional in-film polarization (homogeneous in the domain wall plane), which leads to the Néel domain wall deformation once the tip electrode potential is applied. Due to existing of the another type (in-plane) of the polarization component, more complex (less symmetric) domain wall bend appear in contrast to simple shift. In order to qualitatively describe the polarization components in general case, the detailed classification of the FME coupling in all possible magnetic domain walls (including 0° type) can be utilized [51]. These ($\tilde{\gamma}_8$ and $\tilde{\gamma}_{14}$) FME invariants are possible in the crystals with tetragonal cyclic crystallographic point groups, e.g. cyano-bridged FeNb ferromagnetic material with space group $I4/m1'$ in the paramagnetic phase [52]. The indirect approach to such experimental investigations can be seen in the enhancement of electromagnetooptical effect (i.e. electric field induced Faraday rotation) in the vicinity of domain wall observed in the yttrium-ferrite-garnet films [27].

Point groups of the hexagonal and trigonal crystal systems are either cyclic or dihedral. Consequently, the structure of FME coupling free energy expression is similar in case of all uniaxial crystals. All these expressions have special role of $z$-th components (table 1). Therefore, the above-mentioned effects (described by $\tilde{\gamma}_8$ and $\tilde{\gamma}_{14}$ constants) are expected in all optically uniaxial crystals. Oppositely, the optically isotropic (cubic) crystals have not these effects.

Chiral structures (e.g. vortices and skyrmions) can be induced by the FME effect in any magnetically ordered media. Collapse of the magnetic skyrmion to the atomic scale [53] requires space group based microscopical analysis (similar as [6]) of the magnetoelectric coupling. Skyrmions can be generated by any type of the DM interaction, which conforms to the FME coupling or native interaction of the chiral medium. In the latter case, the DM vector is determined by the crystal axes. In the former case, the DM vector can be expressed via the polarization and $\mathbf{e}_{12}$ on the symmetry based ground. As far as the magnetic moment is the axial time-odd vector, the product $[\mathbf{S}_1 \times \mathbf{S}_2]$ is the axial time-even vector. The product $[\mathbf{e}_{12} \times [\mathbf{S}_1 \times \mathbf{S}_2]]$ is, therefore, time-even polar vector. Finally, the free energy invariant is $\mathbf{P}[\mathbf{e}_{12} \times [\mathbf{S}_1 \times \mathbf{S}_2]]$, which exist in crystal of any symmetry. This conclusion conforms to the similar conclusion in Ref. [29]. The polarization (6) is derived from this invariant, taking into account polarization inducing theory described in the Ref [14]. However, the triad $M - O - M$ is part of the crystal lattice, so its physical properties must satisfy the space group symmetry requirements, which gives the corresponding properties of the tensor $\hat{d}$. Microscopic magnetoelectric coupling details are averaged, and final macroscopic expression (4) is determined by the crystallographic point group $G_P$ [13]. This is the inhomogeneous contribution of the DM interaction to the macroscopic free energy density [30,44]. Anisotropic terms (including antisymmetric ones) retain their form after transition to the macroscopic description. Next in order of magnitude term of the expression (7) is the anisotropic symmetric superexchange interaction (pseudodipolar type) $\mathbf{S_1}\hat{\Gamma}\mathbf{S_2}$ [42]. Its type "relativistic" means that it is quadratic against spin-orbit term in the series of the superexchange energy [42].

## 5. Conclusions

Thus, the phenomenological theory of the flexomagnetoelectric coupling in the any symmetry crystals of the cubic, tetragonal and orthorhombic families has been suggested. The significant role of the crystallographic point group types (symmetric, alternating, dihedral and cyclic) in the flexomagnetoelectric coupling has been shown. The tetragonal crystals with cyclic crystallographic point groups ($4_z/m_z$, $\bar{4}_z$ and $4_z$) have most complex flexomagnetoelectric properties, which are characterized by the 21 phenomenology constants. Six different types of the Lifshitz invariants are required by the symmetry in these crystals.

Verification of the described phenomenological theory can be provided in the experiments with the magnetic domain wall shift / bend. Symmetry and parity of such effect are derived directly from the expressions specified in the present work for any crystal of the cubic, tetragonal and orthorhombic crystal systems. Particularly, the verification of the inhomogeneous magnetization magnitude terms can be provided in the vicinity of the compensation point. In this case, strong enough electric field can induce the changes of the magnetization magnitude in the local region, e.g. in the volume of the magnetic domain wall near the tip electrode.

## Acknowledgements

I would like to express my sincere gratitude to Acad. V.G. Bar'yakhtar, who inspired me to the present work. I thank Prof. V.F. Kovalenko, Prof. V. A. L'vov, Dr. S. V. Olszewski, and Dr. O. V. Tychko for their helpful discussion and suggestions.

**References**

[1] W. Eerenstein, N. D. Mathur, J. F. Scott, Nature 442 (2006) 759-765.

[2] M. Bibes and A. Barthelemy, Nature Materials 7 (2008) 425.

[3] D. Khomskii, Physics 2 (2009) 20.

[4] L. Landay, E. Lifshitz, Electrodynamics of Continuous Media, Pergamon Press, Oxford, 1965.

[5] G. Smolenskii and I. Chupis, Uspehi Fiz. Nauk 137 (1982) 415.

[6] I. Dzyaloshinskii, Zh. Eksp. Teor. Fiz., 37 (1959) 881.

[7] N. Neronova and N. Belov, Dokl. Akad. Nauk SSSR 120 (1959) 556.

[8] V.G. Bar'yakhtar, V.A. L'vov, D.A. Yablonskiy, JETP Lett. 37, 12 (1983) 673.

[9] A. A. Khalfina and M. A. Shamtsutdinov, Ferroelectrics 279 (2002) 19.

[10] G. A. Smolenskii, I. Chupis, Sov. Phys. Usp. 25(7) (1982) 475.

[11] I. M. Vitebskii, D. A. Yablonski, Sov. Phys. Solid State 24 (1982) 1435.

[12] A. Sparavigna, A. Strigazzi, A. Zvezdin, Phys. Rev. B 50 (1994) 2953.

[13] V.G. Bar'yakhtar, V.A. L'vov, D.A. Yablonskiy, Theory of electric polarization of domain boundaries in magnetically ordered crystals, in: A. M. Prokhorov, A. S. Prokhorov (Eds.), Problems in solid-state physics, Chapter 2, Mir Publishers, Moscow, 1984, pp. 56-80.

[14] M. Mostovoy, Phys. Rev. Lett. 96 (2006) 067601.

[15] A. P. Pyatakov, G. A. Meshkov, PIERS Proceedings, Cambridge, USA, July 5-8, 2010

[16] A.P. Pyatakov, A.K. Zvezdin, Eur. Phys. J. B 71 (2009) 419.

[17] A. K. Zvezdin, A. A. Mukhin, JETP letters 89, 7 (2009) 328.

[18] A.P. Pyatakov, A.K. Zvezdin, physica status solidi (b) 246, 8 (2009) 1956.

[19] Z.V. Gareeva, A.K. Zvezdin, Phys. Sol. St. 52, 8 (2010) 1714.

[20] A.P. Pyatakov, A.K. Zvezdin, Phys.-Usp. 52 (2009) 845.

[21] A.P. Pyatakov, A.K. Zvezdin, Low Temp. Phys. 36 (2010) 532.

[22] Z.V. Gareeva, A.K. Zvezdin, Phys. Stat. Sol. (RRL) 3 (2-3) (2009) 79.

[23] B.M. Tanygin, IOP Conf. Ser.: Mater. Sci. Eng. 15 (2010) 012073.

[24] Stefanovskii et al., Sov. J. Low Temp. Phys. 12 (1986) 478.

[25] A. S. Logginov et al., JETP Lett. 86 (2007) 115;
   Appl. Phys. Lett. 93 (2008) 182510.

[26] A.P. Pyatakov, et al., Europhys. Lett. 93 (2011) 17001.

[27] V.E. Koronovskyy, S.M. Ryabchenko, V.F. Kovalenko, Phys. Rev. B 71 (2005) 172402.

[28] B.M. Tanygin, J. Magn. Magn. Mater 323 (2011) 1899.

[29] I.E. Dzyaloshinskii, Europhys. Lett. 83(6) (2008) 67001.

[30] I.E. Dzyaloshinskii, Sov. Phys. JETP 19 (1964) 960.

[31] I.E. Dzyaloshinskii, Sov. Phys. JETP 10 (1960) 628.

[32] G. J. Legg and P. C. Lanchester, J. Phys. C: Solid State Phys. 13 (1980) 6547.

[33] Y.-D. Li, K. Y. Lee, and F.-X. Feng, Int. J. Sol. Struct. 48 (9) (2011) 1311.

[34] H. T. Chen, L. Hong, and A. K. Soh, J. Appl. Phys. 109 (2011) 094102.

[35] J. X. Zhang, Y. L. Li, D. G. Schlom, and L. Q. Chen, Appl. Phys. Lett. 90 (2007) 052909.

[36] J.J. Betouras, G. Giovannetti, and J. Brink, Phys. Rev. Lett. 98 (2007) 257602.

[37] H. Katsura, N. Nagaosa, and A.V. Balatsky, Phys. Rev.Lett. 95 (2005) 057205.

[38] K.T. Delaney, M. Mostovoy and N.A. Spaldin, Phys. Rev.Lett.102 (2009) 157203.

[39] C. D. Hu, Phys. Rev. B 77 (2008) 174418.

[40] I. A. Sergienko and E. Dagotto, Phys. Rev. B 73 (2006) 094434.

[41] I. Dzyaloshinskii, J. Phys. Chem. Solids 4 (1958) 241.

[42] T. Moriya, Phys. Rev. 120 (1960) 91.

[43] T. Moriya, J. Appl. Phys. 39 (1968) 1042.

[44] A.N. Bogdanov et al., Phys. Rev. B 66 (2002) 214410.

[45] K.W.H. Stevens, Revs. Modern Phys. 25 (1953) 166.

[46] M. Mostovoy, Phys. Rev. Lett. 94 (2005) 137205.

[47] M. Mostovoy, K. Nomura, and N. Nagaosa, Phys. Rev. Lett. 106 (2011) 047204.

[48] P. Chandra, and P. B. Littlewood, Ferroelectr. 105 (2007) 69.

[49] T. W. Hungerford, Algebra, Volume 73 of Graduate Texts in Mathematics, Springer, 1980.

[50] B.M. Tanygin, Physica B 406 (2011) 3423.

[51] B.M. Tanygin, J. Magn. Magn. Mater 323 (5) (2011) 616.

[52] D. Pinkowicz et al., Dalton Trans. 37 (2009) 7771.

[53] A. Abanov and V. L. Pokrovsky, Phys. Rev. B 58 (1998) R8889.